\documentclass{elsarticle}
\usepackage[utf8]{inputenc}    
\usepackage{graphicx}          
\usepackage{comment}           
\usepackage{indentfirst}       
\usepackage{amsmath}           
\usepackage{array, threeparttable} 
\usepackage{multirow}          
\usepackage[numbers]{natbib}
\usepackage{booktabs}          
\usepackage[font=small,labelfont=bf,labelsep=none]{caption} 
\usepackage{geometry}          
\usepackage{lineno}            
\usepackage{hyperref}          

\begin{document}
\title{A Survey of Foundation Models for Music Understanding}

\author[a]{Wenjun Li\fnref{equal}}
\author[b]{Ying Cai\fnref{equal}}
\author[a]{Ziyang Wu\fnref{equal}}
\author[a]{Wenyi Zhang}
\author[a]{Yifan Chen}
\author[a]{Rundong Qi}
\author[a]{\\Mengqi Dong}
\author[a]{Peigen Chen}
\author[a]{Xiao Dong}
\author[a]{Fenghao Shi}
\author[a]{Lei Guo}
\author[a]{Junwei Han}
\author[b]{\\Bao Ge}
\author[c]{Tianming Liu\corref{corresponding}}
\ead{tliu@cs.uga.edu}
\author[a]{Lin Gan\corref{corresponding}}
\ead{ganlin@nwpu.edu.cn}
\author[a]{Tuo Zhang\corref{corresponding}}
\ead{tuozhang@nwpu.edu.cn}

\cortext[corresponding]{Corresponding author.}
\fntext[equal]{The three authors contribute equally to this work.}

\affiliation[a]{
organization={School of Automation, Northwestern Polytechnical University},
city={Xi'an},
postcode={710129},
state={Shaanxi},
country={China}
}
\affiliation[b]{
organization={School of Physics and Information Technology, Shaanxi Normal University},
city={ Xi'an},
postcode={710119},
state={Shaanxi},
country={China}
}

\affiliation[c]{
organization={School of Computing, The University of Georgia},
city={Athens},
postcode={30602},
country={USA}
}

\date{June 2024}

\begin{abstract}
Music is essential in daily life, fulfilling emotional and entertainment needs, and connecting us personally, socially, and culturally. A better understanding of music can enhance our emotions, cognitive skills, and cultural connections. The rapid advancement of artificial intelligence (AI) has introduced new ways to analyze music, aiming to replicate human understanding of music and provide related services. While the traditional models focused on audio features and simple tasks, the recent development of large language models (LLMs) and foundation models (FMs), which excel in various fields by integrating semantic information and demonstrating strong reasoning abilities, could capture complex musical features and patterns, integrate music with language and incorporate rich musical, emotional and psychological knowledge. Therefore, they have the potential in handling complex music understanding tasks from a semantic perspective, producing outputs closer to human perception. This work, to our best knowledge, is one of the early reviews of the intersection of AI techniques and music understanding. We investigated, analyzed, and tested recent large-scale music foundation models in respect of their music comprehension abilities. We also discussed their limitations and proposed possible future directions, offering insights for researchers in this field.
\end{abstract}

\maketitle

\section{Introduction}
Music is intricately woven into people's daily life, meeting both emotion and entertainment needs. Its significance is clear not only on a personal level but also within social and cultural contexts, making it an essential aspect of human existence, culture, and society. In essence, efforts in understanding music could help deepen our emotional experiences, enhance cognitive abilities and 
enrich our lives, and further connect us with others and diverse cultures
\cite{girdhar2023imagebind, herndon1981music, bohlman2003music, cross2001music, juslin2001music, lepa2020popular, rentfrow2012role}. Nowadays, the rapid development of AI provides a new and exciting avenue to music understanding\cite{davies2011musical,zeng2021musicbert}. A variety of techniques and models have been developed by researchers to fulfill tasks including music theory analysis, music information retrieval, music perception and cognition, performance analysis\cite{dunsby1988music,casey2008content}, attempting to enable machines to understand music as intricately and authentically as humans do and provide services that people need, such as music creation, analytical research, music education, and emotional recognition in music\cite{holland2013artificial}.


With the rapid development and breakthrough of deep learning models, such as Convolutional Neural Networks (CNNs)\cite{alzubaidi2021review}, Generative Adversarial Networks (GANs)\cite{goodfellow2020generative}, Recurrent Neural Networks (RNNs)\cite{medsker2001recurrent}, and Transformers, significant success in the fields of computer vision and natural language processing (NLP) has been achieved. This success includes applications such as machine translation, speech recognition, and text generation, fundamentally transforming our way of life \cite{liu2023summary}. Meanwhile, music, often being referred as a universal language \cite{mcmullen2004music}, shares similarities with natural language in terms of both format as input to machine and their abstract functions in emotional and conceptional expression. Therefore, researchers have become interested in applying these aforementioned deep learning models to the field of music. For instance, in the realm of music understanding, these earlier technologies were adept at handling basic tasks, including music classification and generating simple music captions. Below, we present some examples of such early AI methods.


Thomas Pellegrini and colleagues\footnote{ \url{https://github.com/topel/audioset-convnext-inf}}\cite{pellegrini2023adapting} proposed applying Depthwise Separable Convolution (DSC)\cite{chollet2017xception} to the Pretrained Audio Neural Network (PANN)\cite{kong2020panns} family for audio classification on AudioSet to show its advantages in accuracy and model size. This model performs well in music classification tasks. The Audio Spectrogram Transformer (AST)\footnote{\url{https://github.com/YuanGongND/ast}}\cite{gong2021ast}, a convolution-free, purely attention-based model for audio classification which features a simple architecture and superior performance. This model is applied directly to audio spectrograms. It can capture long-term global context even at the lowest layer. MusCaps\footnote{\url{https://github.com/ilaria-manco/muscaps}}\cite{manco2021muscaps} combines CNN and RNN to generate music captions through a pre-trained audio encoder.

%

In spite of their achievement, these music understanding models are only feasible in certain situations, particularly when computational resources are limited or the tasks are relatively simple. One possible reason is that the way these models process music is significantly different from humans. These models lack an understanding of musical internal logic, such as tonal, chord features and the coherent structure, as well as the context where music is performed. They are only trained on the audio features of music and use one-hot encoding to assist with various simple and single-type tasks, such as classification\cite{fu2010survey}, resulting in an inability to integrate musical ontological structural features beyond the audio features \cite{ALSHOSHAN200695}. Aside from single-type tasks, they are unable to handle cross-domain or personalized tasks based on specific instructions. Moreover, due to the limited parameters of these models, their performance in complex tasks is not satisfied and lacks creativity. Strictly speaking, these music models can not complete music understanding tasks but perform downstream tasks simply on audio signals.

By virtue of advancements in graphics processing unit chips (GPUs), artificial intelligence has progressed significantly with the development of large language models (LLMs) and foundation models (FMs). These models are pre-trained on extensive text, image, and audio data and have a vast number of parameters, showcasing their potential across various fields such as text-based question answering, text summarization, image generation, audio generation, image analysis, and audiovisual recognition\cite{han2021pre}. 
In this sense, numerous parameters of large foundation models enable them to capture complex musical features and patterns, and to integrate music features with language through cross-modal feature alignment and fusion\cite{zhang2021heterogeneous}. They can integrate semantic information and possess strong reasoning abilities\cite{huang2022towards,wei2022chain}, addressing many limitations of basic models in music understanding. Leveraging their strong human-like reasoning and contextual understanding capabilities, these models can tackle complex music understanding tasks from a semantic perspective rather than merely signal classification. Also, large foundation models make it possible to incorporate knowledge and feedback of human musicians from different regions, times and fields with various evaluation perspectives, and to yield output closer to human \cite{kang2024self}. Being trained on professional corpora and high-quality domain-specific texts, large foundation models have rich emotional analysis\cite{yang2024advancing} and psychological knowledge\cite{pellert2023ai,lai2023psy,rathje2024gpt}, enhancing their ability to understand and express emotions in music. Many researchers have utilized this technology to develop models with music understanding capabilities, achieving notable results in tasks such as multi-label music classification and music scenario reasoning\cite{gardner2023llark}. In a word, although current large music models still have shortcomings in some aspects, such as associating subtle emotional recognition with the internal logical structure of music, they overcome many limitations faced by previous basic models and demonstrate the enormous potential of large foundation models technology in the field of music understanding.

By integrating perspectives from AI practitioners and music experts, this paper aims to provide a comprehensive overview of how AI and music intersect, focusing on the development and application of large music models, particularly in music understanding domain.
We have reviewed various large music models and multimodal large language models, offering insights into their development, technologies, and applications. Additionally, we have collected and organized music datasets for machine learning, including detailed information on data format, source, label types, and audio duration. Furthermore, recognizing the critical role of evaluation metrics in machine learning\cite{zhou2021evaluating}, which determine the overall effectiveness of the outcomes, we have reviewed relevant literature and consulted music experts to summarize and present the evaluation metrics for music in machine learning, focusing on both subjective and objective evaluations. Furthermore, we have conducted relevant tests on music understanding using existing publicly available models and presented the test results. Finally, we analyzed the results, discussing the reasons for the differences in test outcomes, the limitations, and shortcomings of the existing large music models, and listing some feasible future work. We hope our work could help readers quickly establish their research processes and determine their research directions based on the provided resources.

The following sections are organized as follows: In Section {2}, we summarize and introduce music datasets and evaluation metrics. In Section {3}, we present existing large music models and multimodal large language models applicable to music understanding, providing an initial understanding of the large multimodal model(LMM) for music domain. Following this, in Section {4}, we evaluated the music understanding ability of several models through experiments comparing and summarizing their music perception abilities. Finally, in Section {5}, we discuss significant factors constraining the development of music in artificial intelligence, offering suggestions for the future directions and advancements in computer-based music understanding.

\section{Dataset and Evaluation Metrics}
To develop and implement an artificial intelligence model algorithm, it is essential to first understand the datasets and evaluation metrics relevant to the specific task. Only after having these two elements can we proceed with the actual algorithm development. Therefore, in this section, we will introduce several representative open-source music datasets we have collected, as well as the relevant evaluation metrics in the music understanding domain, and provide detailed explanations.
\subsection{Datasets}
Selecting the appropriate dataset for different music artificial intelligence tasks is crucial, as the quality and type of the dataset will directly impact the effectiveness of the developed algorithms. We have gathered several representative music datasets to provide readers with a comprehensive understanding of music-related datasets. We describe these datasets based on various aspects, including popularity, data format, the duration of every music sample, label type, and the uniqueness of annotations (tags). By gaining an understanding of the existing music datasets, readers can quickly comprehend and choose the suitable dataset according to their specific needs to develop music understanding AI applications.

For clarity, the datasets were grouped or ordered according to the tags they provided in the following sections and Table \ref{tab:my_label}, including genre, music mood, music caption, instruments and mixture of them. Other features of these datasets including file format and duration of samples were also reported.

\begin{table}[!ht]
\centering
    \caption{\ Representative datasets for music understanding.}
    \begin{tabular}{cccc}
    \toprule[2pt] 
         Dataset & Tag & Format & Duration \\
   \midrule[2pt] 
         GTZAN & genre & .wav & 30s \\
         Music Genre & genre & .mp3 & 270s$\sim$300s  \\
         EDM Music Genres & genre & .wav & 3s \\
         FMA & genre & .mp3 & 30s \\
         Music\_Classification & music mood & .wav & 5s \\
         Multi-modal MIREX Emotion & music mood & .mp3\&.txt\&.midi & 30s \\
         CTIS & instrument & .wav & 3s \\
         The Lakh MIDI & beats\&keys\&rhythm & .midi & 180s$\sim$300s \\
         MusicCaps & music caption & .wav & 10s \\
         MTG-Jamendo & genre\&instrument\&mood\&theme & .mp3 & 30s \\
    
    \bottomrule[2pt]
    \end{tabular}
    \label{tab:my_label}
\end{table}

\textbf{Genre}

GTZAN Genre Collection (GTZAN)\footnote{ \url{https://www.kaggle.com/datasets/andradaolteanu/gtzan-dataset-music-genre-classification?select=Data}}\cite{tzanetakis2002manipulation,tzanetakis2002musical} contains 1000 music pieces in .wav format, divided into 10 different music genres: Blues, Classical, Country, Disco, Hip-Hop, Jazz, Metal, Popular, Reggae, and Rock. There are 100 examples of each of these music genres, each of which contains a piece of music with a duration of 30 seconds. The GTZAN dataset has been downloaded up to 3.1k times.

\sloppy
The Music Genre Dataset\footnote{ \url{https://huggingface.co/datasets/ccmusic-database/music_genre}}contains 1700 music works in .mp3 format, which are derived from NetEase Music, divided into nine music genres: Symphony, Opera, Solo, Chamber, Pop, Dance and House, Indie, Soul or R\&B, Rock. The duration of each clip ranges from 270 seconds$\sim$300 seconds.

The EDM Music Genres (EDM)\footnote{ \url{https://www.kaggle.com/datasets/sivadithiyan/edm-music-genres}} is derived from YouTube music clips and contains 40,000 pieces of music which divided into 16 different genres: Ambient, Big Room House, Drum and Bass, Dubstep, Future Garage/Wave Trap, Hardcore, Hardstyle, House, Lo-fi, Moombahton/Reggaeton, Phonk, Psytrance, Synthwave, Techno, Trance, Trap. There are 2,500 examples for each music genre, 2,000 for training features, and 500 for testing features, each containing a music clip with a duration of 3 seconds.

The Free Music Archive (FMA) dataset \footnote{ \url{https://github.com/mdeff/fma}}\cite{defferrard2016fma}  has a large amount of data that provides a diverse sample of more than 100,000 pieces of music in .mp3 formats and is designed to facilitate research in music classification, recommender systems, and other related fields. The FMA dataset contains 161 genres, such as popular, classical, folk, etc. Each of these music samples lasts for 30 seconds.

\textbf{Music mood}

The Music\_Classification dataset \footnote{ \url{https://www.kaggle.com/datasets/shanmukh05/music-classification}} is in .wav format and contains 5 different musical moods: aggressive, dramatic, happy, romantic, sad. There are a total of 10,133 examples, each of which contains a music clip with a duration of 5 seconds.

The Multi-modal MIREX Emotion Dataset \footnote{ \url{https://www.kaggle.com/datasets/imsparsh/multimodal-mirex-emotion-dataset}}\cite{panda2013multi} has three formats, including .mp3, .txt, and .midi, and the same music clip is named in the same way in different formats. This dataset contains audio, midi and lyrics and 5 different musical moods: passionate, rollicking, literate, humorous and aggressive. There are a total of 903 examples, each of which contains a music clip with a duration of 30 seconds.

\textbf{Instruments}

Chinese Traditional Instrument Sound (CTIS) \footnote{ \url{https://huggingface.co/datasets/ccmusic-database/CTIS}}\cite{ZYYX202002013,liang2019constructing,li2018dcmi} is a dataset containing sound information about Chinese traditional musical instruments. It includes 287 Chinese national musical instruments, including traditional musical instruments, improved musical instruments and ethnic minority musical instruments. The music files of dataset is saved by .wav format. It selects the typical sound of all conventional performance techniques, and the representative fragments in music. Meanwhile, it makes comprehensive annotation on the playing position, pitch and performance techniques of musical instruments.The duration of each track is 3 seconds.

\textbf{MIDI annotations}

The Lakh MIDI dataset\footnote{\url{https://colinraffel.com/projects/lmd/}}\cite{raffel2016learning} is a collection of 176,581 unique midi files, using information extracted from midi files as annotations for matching audio files. You can get a transcription of the song, as well as meter information such as beats and keys. The duration of each track ranges from 180 seconds to 300 seconds. The author did not provide the total duration of all the data.

\textbf{Music caption}

MusicCaps \footnote{ \url{https://www.kaggle.com/datasets/googleai/musiccaps}}\cite{agostinelli2023musiclm} is a dataset of 5.5k music-text pairs, containing 5521 music examples in .wav format, each labeled with a list in English (eg.pop, tinny wide hi hats, mellow piano melody, high pitched female vocal melody, sustained pulsating synth lead) and free-text descriptive subtitles written by musicians (e.g., "This folk song features a male voice singing the main melody in an emotional mood. This is accompanied by an accordion playing fills in the background. A violin plays a droning melody. There is no percussion in this song. This song can be played at a Central Asian classical concert.” ). These texts focus on the sound of the music, not metadata. Each instance contains a piece of music with a duration of 10 seconds.

\textbf{Multiple tags}

The MTG-Jamendo dataset\footnote{\url{https://github.com/MTG/mtg-jamendo-dataset}}\cite{bogdanov2019mtg} is an open music auto-tagging dataset built using music available on the Jamendo platform under Creative Commons licenses and tags provided by content uploaders. This dataset contains over 55,000 full-length audio tracks annotated with 195 tags across categories such as genre, instrument, and mood/theme. The dataset offers detailed data splits and benchmarking, with audio tracks encoded in 320kbps mp3, totaling 3,777 hours of music. 

\subsection{Evaluation Metrics}
When developing a music understanding model, we need to clarify the relevant evaluation indicators and metrics, the appropriate selection of which is equally vital. The evaluation system comprises both objective and subjective metrics: objective metrics provide clear, data-based comparisons, while subjective metrics help to gauge the popularity of the music generated by each model among listeners and their assessment of its artistic value. Typically, a comprehensive evaluation system combines these two types of metrics to thoroughly assess and compare the performance of various music models.
\subsubsection{Subjective Evaluation Metrics}
In testing the music comprehension ability of the large music model, subjective evaluation is crucial because it provides direct feedback on the perceived quality, emotional expression, cultural relevance, and creativity of a musical piece, aspects that are often not fully captured by objective metrics. Subjective evaluation of music datasets is typically performed using human listening tests to measure performance on various aspects of the music. The method is used in a similar way to a Turing test or a questionnaire.
The mean opinion score(MOS)\cite{streijl2016mean} is one of the subjective evaluation method that determines the quality of audio by subjective score of human raters, so it is flexible to test different aspects of music. The MOS score range is 0$\sim$5 points that following metrics in a five-point Likert scale. The larger the score, the better the music quality. When evaluating the MusicCaps\cite{agostinelli2023musiclm} dataset, the researchers collected a total of 1,200 ratings involving 600 pairwise comparisons from each source. In the music test, participants were shown two 10-second clips and a text title and were asked which clip best described the title text on a 5-point Likert scale (e.g., “Which of the music clips is best described by the text? A. Strong preference for option 1. B. Weak preference for option 1. C. No preference. D. Strong preference for option 2. E. Weak preference for option 2.") They were also instructed to ignore audio quality and focus only on how well the text fit with the music.

\subsubsection{Objective Evaluation Metrics}

While human rating of music may be a reasonable method of subjective assessment of music, it faces several challenges in machine learning tasks: it is difficult to standardize, results lack consistency due to individual differences, implementation can be costly, and preference may influence outcomes. 
Therefore, it is necessary to add objective evaluation indicators and combine them with subjective evaluation ones.

The objective evaluation of a music understanding system's performance includes assessing the model's accuracy in identifying qualitative indicators such as genre and instruments, as well as its alignment with quantitative indicators based on music rules and the actual music itself. 
Here are metrics used for objective evaluation in the literature.

\textbf{Rhythm}

a)Beat Per Minute(BPM)\cite{orio2006music}: The number of beats per minute, used to measure the speed of music. BPM is an accurate value that objectively indicates the speed of music and is a basic indicator of rhythm.

b)Time Signature\cite{cooper1963rhythmic,lerdahl1996generative}: The denominator indicates the note value that represents one beat, and the numerator indicates the number of beats per measure, such as 4/4, 3/4, 6/8. The time signature clearly defines the basic rhythmic unit and structure of a measure, forming the basis of rhythm.

c)Rhythmic Complexity\cite{vuust2014rhythmic}: The complexity of rhythm, which can reflect the compositional techniques and expressiveness of the music. It is an important metric for music performance and analysis.

\textbf{Key and Harmony}

a)Key\cite{boone2017music}: The tonic and scale of the music, such as C major, A minor, etc. The key defines the pitch system and harmonic relationships of the music, forming the basis for understanding melody and harmony.

b)Harmonic Complexity\cite{huron1994interval,plomp1965tonal,sethares2005tuning}: The complexity of chords in music. Harmonic complexity can reflect the technical difficulty and depth of expression in music, making it an important indicator for music analysis.

\textbf{Instrumentation}

a)Instrumentation: The types of instruments used in music. Different combinations of instruments affect the timbre and expressiveness of the music, making instrumentation a crucial element in musical arrangement.

b)Timbre\cite{asa1960acoustical}: The tonal characteristics of instruments, including frequency and time domain indicators. Timbre is an important feature distinguishing different sounds, affecting the personality and recognizability of music.

c)Number of Voices: The number of independent parts in the music. The number of voices influences the texture and richness of the music, making it an important element in arranging and composing.

\textbf{Genre}

The style of music, such as Rock, Classical, Jazz, etc. Genre is a classification standard for music, helping to understand the historical background and cultural characteristics of the music.

\textbf{Musical Structure}

The sectional structure of music\cite{lerdahl1996generative}, such as A-B-A form. Musical structure determines the overall layout and development of the music, making it a crucial aspect of understanding and analyzing musical works.

\section{Large Foundation Models for Music Understanding}

We have collected several large foundation models that have been released in the past years and are capable of understanding music. As shown in Table \ref{Table:character}.

\begin{table}[h!]
\centering
\caption{\ Model Characteristics. Music specific models and general models are separated.I stands for image, M stands for music, A stands for audio and T stands for text.}
\begin{tabular}{lcccc}
\toprule
Model & Language Model & Parameters & Input Modalities & Output Modalities \\
\midrule
Qwen-Audio & Qwen-7B & 7.7 B & M, A, T & M, A, T \\
LTU & LLaMA-7B & 7 B & M, A, T & T \\
SALMONN & Vicuna LLM & 7 B &M, A, T & T\\
ModaVerse & Vicuna LLM & 7 B & I, M, A, T & 
 I, M, A, T \\
AnyGPT & LLaMA 2-7B & 8 B & I, M, A, T & I, M, A, T \\
\cmidrule(lr){1-5}
ChatMusician & LLaMA 2-7B & 7 B & M, T & M, T \\
M²UGen & LLaMA 2-7B & 7 B & I, M, T & M, T \\
MU-LLaMA & LLaMA 2-7B & 7 B & M, T & T \\

\bottomrule
\end{tabular}

\label{Table:character}

\end{table}

\begin{figure}[!ht]
    \centering
    \includegraphics[width=16cm]{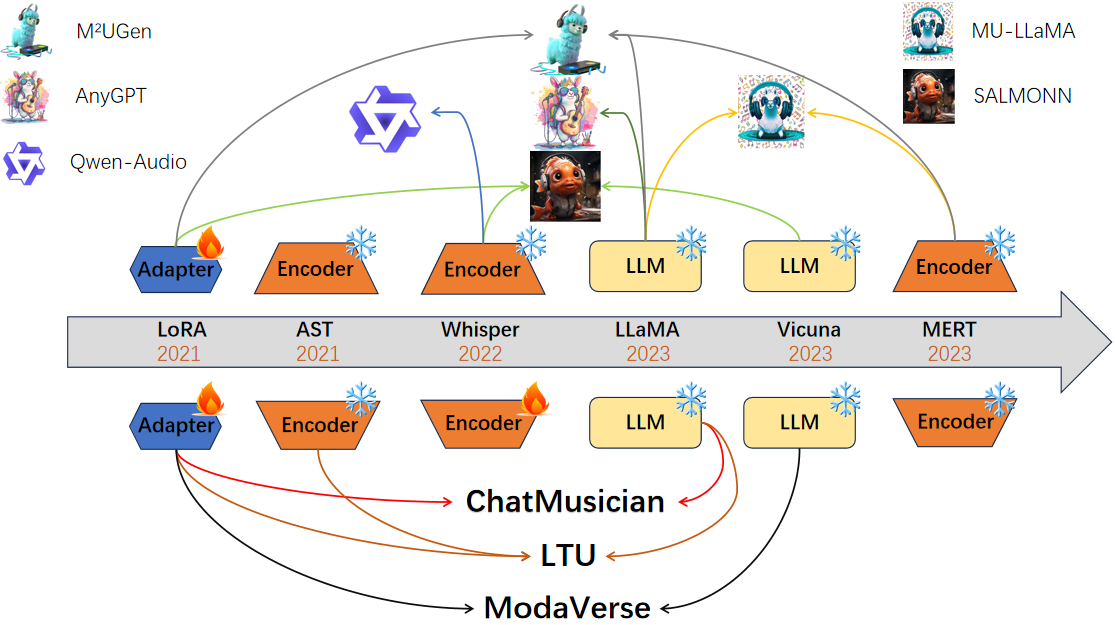}
    \caption{\ Related technologies for the model architecture of the music understanding module in the music large foundation model. The flame icon represents trainable, while the snowflake icon represents frozen state.}
    \label{fig:architectures}
\end{figure}

These models are based on large language models and have been extended and improved for specific modal tasks. Each model has different key modules, such as music encoders, music decoders, and the intermediate large language models, each with its own unique characteristics. The training strategies adopted by different models also vary, leading to differences in their focus areas and results in music understanding. In Figure \ref{fig:architectures}, we illustrate representative technologies of the large music model architecture over time, including the large language model module, music encoder module, and adapter module. In Table \ref{Table:character}, we have organized the relevant parameters of the model, including model name, language model module, model parameters, input modality, and output modality. These models we collected are divided into two categories: 1) large foundation models that specifically focus on music and possess both music understanding and generation capabilities; 2) general multimodal large foundation models with music understanding capabilities. The modules used in each model illustrated in Figure \ref{fig:architectures} are discussed in details in the following sections.

\subsection{Music Specialized LLMs}

\subsubsection{ChatMusician}

ChatMusician\cite{yuan2024chatmusician} is a large language model that combines music comprehension and generation. Its design is based on the LLaMA 2\cite{touvron2023llama} model with continuous pre-training and fine-tuning. To enable music understanding and generation, ChatMusician employs an ABC note representation which is compatible with text. Unlike traditional methods, this model does not require external multimodal neural architectures or dedicated taggers. With this approach, ChatMusician is able to understand and generate diverse musical content including text, chords, melodies, themes, and musical forms, outperforming baseline models such as GPT-4\cite{achiam2023gpt}.

ChatMusician uses a pre-training dataset called MusicPile, which is specifically designed to inject musical capabilities into LLMs. In addition, metadata of 2 million music tracks were crawled from YouTube and associated question-answer pairs were generated. During training, the model adopts continuous pre-training and fine-tuning, initializes LLaMA 2-7B-Base with fp16 accuracy, and integrates LoRA adapter\cite{hu2021lora}. In training, the ratio of music score to music knowledge and summary data is determined to be 2:1, which ensures the efficiency of music generation and understanding.

In terms of music understanding, ChatMusician demonstrates significant advantages and superior performance. By converting music from .wav or .mp3 files to ABC notation, music can be represented in a highly compressed format, effectively shortening the length of the music sequence and improving the model's processing efficiency. Additionally, this method avoids quantization issues, ensuring that the generated music maintains rhythmic accuracy. Furthermore, the high compatibility of ABC notation with language models enables ChatMusician to perform advanced music analysis in large language model applications.

\subsubsection{M²UGen}
M²UGen\cite{hussain2023m} is a multimodal framework that integrates large language models to understand and generate music across different modalities. This framework relies on pre-trained MERT\cite{li2023mert}, ViT\cite{dosovitskiy2020image}, ViViT\cite{arnab2021vivit} to support comprehensive understanding and generation of music, images, and videos, respectively, and LLaMA 2\cite{touvron2023llama} model to serve as the core large language model, collaborates with these encoders through a multimodal understanding adapter, enhancing capabilities in music comprehension, question answering, and generation. In particular, MERT is used as the music encoder due to its superior performance in music annotation tasks. The framework also employs specialized audio tokens, denoted as [\texttt{AUD}$_i$] where $i \in \{1, 2, \ldots, K\}$ (with $K$ as a hyperparameter representing the number of special audio tokens added to the LLaMA 2 model's vocabulary), to distinguish between music question-answering and generation tasks, ensuring the model can correctly produce either audio or plain text outputs.

The training strategy involves freezing the pre-trained encoders and the generation model, focusing on training the multimodal understanding adapter and the output projector to reduce computational load and cost. The LLaMA 2 model is trained using the LoRA\cite{hu2021lora} method, which simplifies the training process and minimizes the number of trainable parameters. For data generation, the framework utilizes MU-LLaMA\cite{liu2024music} and MPT-7B(MosaicPretrainedTransformer)\cite{mosaicml2023introducing} models to create a multimodal-music paired dataset, addressing the scarcity of music-related task datasets. The generated MUCaps\cite{hussain2023m} dataset is used to align the encoders and decoders, ensuring effective multimodal understanding.

By employing the high-performing MERT model as the music encoder, the M²UGen framework ensures high-quality music input feature embeddings. Additionally, with the use of specialized audio tokens, M²UGen can flexibly produce plain text outputs for music question-answering and audio outputs for music generation within the same framework, catering to different task requirements. This versatility enables the model to excel in music understanding tasks.
\subsubsection{MU-LLaMA}
MU-LLaMA\cite{liu2024music} is a multimodal model that utilizes large language models (LLMs) to answer music-related questions and generate music captions. The model uses pre-trained MERT and LLaMA models for encoding music features and answering music-related questions, respectively. The model also proposes a music understanding adapter to fuse features and feed them into LLaMA. In addition, to improve the model's question-answering capabilities, the model proposes a systematic approach to creating music question-answer datasets, which are essential for training the MU-LLaMA model.

The training strategy involves freezing the pre-trained encoders and the generation model, focusing on training the music understanding adapter to reduce computational load. The LLaMA model is the top 19 layers of the LLaMA 2-7B model. The LLaMA 2 model was fine tuned using the LoRA method.

For data generation, the framework utilizes MPT model to create a multimodal-music paired dataset, addressing the scarcity of question-answering task datasets. The generated MusicQA dataset\cite{liu2024music} is designed to answer open-ended questions related to music.

By employing the high-performing MERT model as the music encoder, the MU-LLaMA model ensures high-quality music input feature embeddings. Additionally, with the use of MPT, MU-LLaMA can generate music question-answer pairs for training the model. This design enables the model to excel in question-answering tasks.

\subsection{General Multimodal Large Language Models}
\subsubsection{Qwen-Audio}
Qwen-Audio\footnote{\url{https://huggingface.co/Qwen/Qwen-Audio}}\cite{chu2023qwen} contains an audio encoder and a large language model. It is a multi-task audio large language model based on audio and text input. Qwen-Audio can provide a variety of audio (including speaker's voice, natural sounds, music, songs) and text as input, and use text as output. It can host more than 30 different audio tasks for training.

Qwen-Audio is initialized with Whisper-large-v2\cite{radford2023robust} as the audio encoder, capable of processing various types of audio, such as human speech, natural sounds, music, and songs. The encoder consists of 640M parameters, which is able to convert the original audio waveform into an 80-channel mel-spectrogram, and reduce the length of the audio representation through the pooling layer, so that each frame output by the encoder corresponds to approximately a 40 millisecond segment of the original audio signal. In addition, Qwen-Audio uses the Qwen-7B pre-training model as the initialization of the language model.

To overcome one-to-many interference, Qwen-Audio uses a wide range of audio datasets for collaborative training, designing a multi-task training framework to encourage knowledge sharing by regulating a series of hierarchical labels of the decoder, and through shared and specified Label separately to avoid interference.

\subsubsection{LTU}

 LTU\cite{gong2023listen} is a multimodal large language model that focus on audio understanding. The model consists of four components, Audio Spectrogram Transformer (AST)\cite{gong2021ast} for encoding music, Audio Projection Layer (APL) for reshaping the feature representation, LLaMA for performing audio understanding tasks, and low-rank adapters (LoRA) for indirectly fine-tuning LLaMA parameters.
In particular, LLaMA is pre-trained on a combination of a natural language corpus and a programming language corpus. In addition, to accommodate a variety of open-ended audio understanding tasks, LTU creates the OpenAQA-5M dataset, which consists of eight mainstream audio datasets and all of the data are (audio, question, answer) tuples.

In order for LTU to learn to answer free-form, open-ended questions based on the given audio, rather than just using the linguistic capabilities of LLaMA, the model divides the training into four phases. In the first phase, LTU freezes the parameters of AST, LoRA and LLaMA, then updates the parameters of the audio projection layer only with the closed-form classification task and the acoustic feature description task.
In the second phase, LTU sets the parameters of all parts except LLaMA as trainable and updates the parameters by running the training tasks from the first phase. The third and fourth phases are similar to the second, except that the third phase extends the training tasks to all closed tasks and the fourth phase extends the training tasks to all closed and open tasks. This design of gradually increasing the difficulty of the training tasks helps the model to perceive and understand the audio, which helps improve the model's performance on open-ended tasks.

Compared to other models, LTU has two advantages when dealing with music-related tasks, firstly, it has high generalisation with the help of OpenAQA-5M dataset, and secondly, it has strong audio perception and comprehension with the help of four-stage training.

\subsubsection{SALMONN}
SALMONN \cite{tang2023salmonn} is a multimodal model designed to understand speech, audio events, and music. It uses a dual encoder architecture with a Whisper speech encoder\cite{radford2023robust} and a BEATs audio encoder\cite{chen2022beats}.Its language model module uses Vicuna LLM\cite{alayrac2022flamingo}, which is a LLaMA LLM fine-tuned to follow instructions. A window-level Q-Former integrates outputs from both encoders into augmented audio tokens, improving its ability to process diverse audio inputs.

SALMONN uses a three-stage cross-modal training method. A large number of tasks  that contain key auditory information but do not require complex reasoning and understanding is used for pre-training, which enables SALMONN to learn high-quality alignment between auditory and textual information.

 In addition, thanks to the Activation Tuning Stage, SALMONN has significant emergent abilities.

\subsubsection{AnyGPT}
AnyGPT \footnote{ \url{https://junzhan2000.github.io/AnyGPT.github.io}}\cite{zhan2024anygpt} builds a text-centric multi-modal alignment dataset AnyInstruct-108k. The dataset consists of 108K multi-turn dialogue samples, which are intricately intertwined with various modes. It aims to use text as a bridge to achieve all the mutual alignment between modalities and enables the model to handle any combination of multimodal inputs and outputs. It is an any-to-any multi-channel language model which can understand and generate a variety of channels, including speech, text, images and music.

AnyGPT employs encoder (a convolutional autoencoder with a latent space quantized using residual vector quantization (RVQ)). This variant processes 32kHz mono audio and achieves a 50Hz frame rate. The generated embeddings are quantized using RVQ with four quantizers, each with a codebook size of 2048, resulting in a combined music vocabulary size of 8192. AnyGPT uses LLaMA as its language model module. In order to enable the language model to predict the entire music piece, the 4-layer music code is flattened into a causal sequence in a frame-by-frame manner. The language model starts by predicting the first four tokens of the first frame and continues predicting subsequent frames in a similar manner.

The field of arbitrary multi-channel large language models (LLMs) is an emerging research field. Therefore, AnyGPT lacks a dedicated benchmark to evaluate the model's capabilities in multiple dimensions, as well as to mitigate potential risks. Although multi-modal LLM with discrete representation can be trained stably, there is a higher loss compared to single-modal training, thus hindering the best performance in each mode. Moreover, AnyGPT limits the music duration to 5 seconds, which greatly limits the practical value of its audio output.

\subsubsection{ModaVerse}

ModaVerse\footnote{\url{https://github.com/xinke-wang/ModaVerse}}\cite{wang2024modaverse} is a multimodal (including audio, image and video) large language model. It employs an Adaptor+Agent training strategy, aligning input features to the language model's textual space through linear projection layers while using existing text-to-X (audio, image, video) models to generate non-text outputs. For musical modality, ModaVerse utilizes a unified encoder, ImageBind\cite{girdhar2023imagebind}, to process various types of inputs, which are then transformed into textual features the model can understand through linear projection layers.ModaVerse uses Vicuna LLM and LoRA fine-tuning technique to reduce training costs and achieve corresponding fine-tuning effects.

Additionally, ModaVerse avoids complex feature-level alignment by directly operating at the natural language level. It completes model training in a single phase through instruction-following tuning, reducing the need for multi-stage training and thus improving training efficiency.

In terms of audio input processing, ModaVerse not only converts audio into textual descriptions but also understands and generates related images or videos. For instance, it can transform the sound of an animal into an image of that animal. This multimodal comprehension capability stems from the efficient alignment of different modal data at the model's input stage, enabling ModaVerse to perform well in music understanding.

\section{Experiments and Analysis}
\subsection{Implementation Details}
We conducted tests on eight existing open-source large foundation models to evaluate their music understanding capabilities, covering both simple and complex music understanding tasks. Due to AnyGPT's inability to perform specific prompts, it was only tested for Music Caption according to the official method.

For simple music understanding tasks, we employed music classification methods. We randomly selected 100 samples from each of the four datasets: GTZAN, Free Music Archive, Music\_Classification, and Multi-modal MIREX Emotion, totaling 400 samples. These samples were then input into the models using a standardized prompt, and the models' responses were recorded. 

For advanced music understanding tasks, we used the MusicCaps dataset. This dataset features professional music evaluations provided by music experts, including sound quality, instruments and performance, emotions, and suitable contexts, making it a high-quality labeled dataset. We randomly selected 147 samples from this dataset, input the test samples along with standardized prompts into the models, and obtained the corresponding captions. These responses were then compared with the reference texts of the actual labels using rouge scores to evaluate the models' advanced music understanding capabilities.
Classification and text similarity metrics are as follows:

\textbf{Classification Metrics}

We use accuracy to characterize the ability of the model for music classification. The reason why we do not use precision and recall is that in some cases, the response of the model may not belong to any of the data labels. But this situation cannot be abandoned, as it also reflects to some extent the ability of the model.

\textbf{Accuracy}

Accuracy is the proportion of correctly classified samples to the total number of samples. As shown in equations (1):

\begin{equation}
\text{Accuracy} = \frac{\text{Number of Correctly Classified Samples}}{\text{Total Number of Samples}}
\end{equation}




\textbf{Text similarity Metrics}

For the text similarity metrics, we use the ROUGE, because the ROUGE has multiple variants such as ROUGE-N and ROUGE-L, which can measure the text similarity at different granularity. This multidimensional method enables the ROUGE to comprehensively reflect the similarity of the text at different levels such as words, phrases and subsequences.

\textbf{ROUGE}

ROUGE (Recall-Oriented Understudy for Gisting Evaluation)\cite{lin2004rouge} is a set of metrics used to evaluate the quality of summaries and translations in natural language processing. ROUGE measures the overlap between the produced texts and a set of reference texts. 

ROUGE-N: Measures n-gram overlap. It calculates the number of matching n-grams between the candidate and the reference texts.As shown in equations (2).

\begin{equation}
\text{ROUGE-N} = \frac{\sum_{\text{S} \in \text{References}} \sum_{\text{gram}_n \in \text{S}} \text{Count}_{\text{match}}(\text{gram}_n)}{\sum_{\text{S} \in \text{References}} \sum_{\text{gram}_n \in \text{S}} \text{Count}(\text{gram}_n)}
\end{equation}

\begin{itemize}
    \item \(\text{References}\): \footnotesize{Set of reference texts}
    \item \(\text{gram}_n\): \footnotesize{n-gram}
    \item \(\text{Count}_{\text{match}}(\text{gram}_n)\): \footnotesize{Number of matching n-grams}
    \item \(\text{Count}(\text{gram}_n)\): \footnotesize{Total number of n-grams in the reference texts}
\end{itemize}

ROUGE-L: Measures the longest common subsequence (LCS). It captures sentence-level structure similarity by measuring the longest matching sequence of words between the candidate and reference texts. As shown in equations (3)-(5).

\begin{equation}
\text{ROUGE-L} = F_1 = \frac{(1 + \beta^2) \cdot \text{Precision} \cdot \text{Recall}}{\beta^2 \cdot \text{Precision} + \text{Recall}}
\end{equation}

\begin{equation}
\text{Precision} = \frac{\text{LCS}(X, Y)}{|X|}
\end{equation}

\begin{equation}
\text{Recall} = \frac{\text{LCS}(X, Y)}{|Y|}
\end{equation}

\begin{itemize}
    \item \(\text{ROUGE-L}\): \footnotesize{ROUGE-L score}
    \item \(F_1\): \footnotesize{F1 score}
    \item \(\beta\): \footnotesize{Weighting factor (usually set to 1)}
    \item \(\text{LCS}(X, Y)\): \footnotesize{Length of the longest common subsequence between X and Y}
    \item \(|X|\): \footnotesize{Length of the candidate summary}
    \item \(|Y|\): \footnotesize{Length of the reference summary}
\end{itemize}





\subsection{Experimental Results}
We conducted tests on eight open-source multimodal large foundation models for music understanding. The downstream tasks performed include genre classification, mood classification, and music captioning.

Due to the fact that objective evaluation metrics based on the music itself, such as rhythm and musical structure, are generally only included in .midi format with relevant labels and not in more common .wav and .mp3 format datasets, we did not perform tasks related to music and rhythm understanding.
For genre classification, we used the GTZAN and FMA datasets. For mood classification, we used the Music\_Classification(MC) and Multi-modal MIREX Emotion(MMME) datasets. Note that, due to the models sometimes providing results outside the given label categories during the identification of music genres, we are unable to calculate metrics such as recall and precision. Therefore, we have analyzed the classification performance solely based on the accuracy of identification. In Table \ref{Table:classification result}, we can observe the performance of various models in Genre and Mood classification tasks. Qwen-Audio excelled in the genre classification task, achieving an accuracy of 80\% and 75\% on the GTZAN and FMA datasets, respectively. However, its performance in mood classification was weaker, particularly on the MC dataset. LTU performed exceptionally well in both genre and mood classification tasks. Its genre classification accuracy on the GTZAN and FMA datasets was 61\% and 70\%, respectively, and its mood classification accuracy on the MC and MMME datasets was 58\% and 79\%, the best among all models. MU-LLaMA performed well in the mood classification task but had average performance in the genre classification task. ModaVerse and M²UGen performed poorly in both tasks.

\captionsetup{justification=centering}
\begin{table}[h!]
\centering
\caption{\ Models Performance on Genre and Mood Classification datasets measured by accuracy. The best performance in each column is highlighted by \pmb{BOLD} type. General models (top section) and music specific models (bottom section) are separated.}
\begin{tabular}{lcccc}
\toprule
Model & \multicolumn{2}{c}{Genre Classification} & \multicolumn{2}{c}{Mood Classification} \\
\cmidrule(lr){2-3} \cmidrule(lr){4-5}
& GTZAN & FMA & MC & MMME \\
\midrule
Qwen-Audio & \pmb{80\%} & \pmb{75\% }& 29\% & 53\% \\
LTU & 61\% & 70\% & \pmb{58\%} & \pmb{79\%} \\
SALMONN & 36\% & 58\% & 22\% & 67\% \\
ModaVerse & 23\% & 44\% & 21\% & 25\% \\    

\cmidrule(lr){1-5}
MU-LLaMA & 51\% & 42\% & 46\% & 74\% \\
ChatMusician & 45\% & 36\% & 39\% & 54\% \\
M²UGen & 7\% & 19\% & 16\% & 41\% \\
\bottomrule
\end{tabular}
\label{Table:classification result}
\end{table}

\begin{figure}[!ht]
    \centering
    \includegraphics[width=16cm]{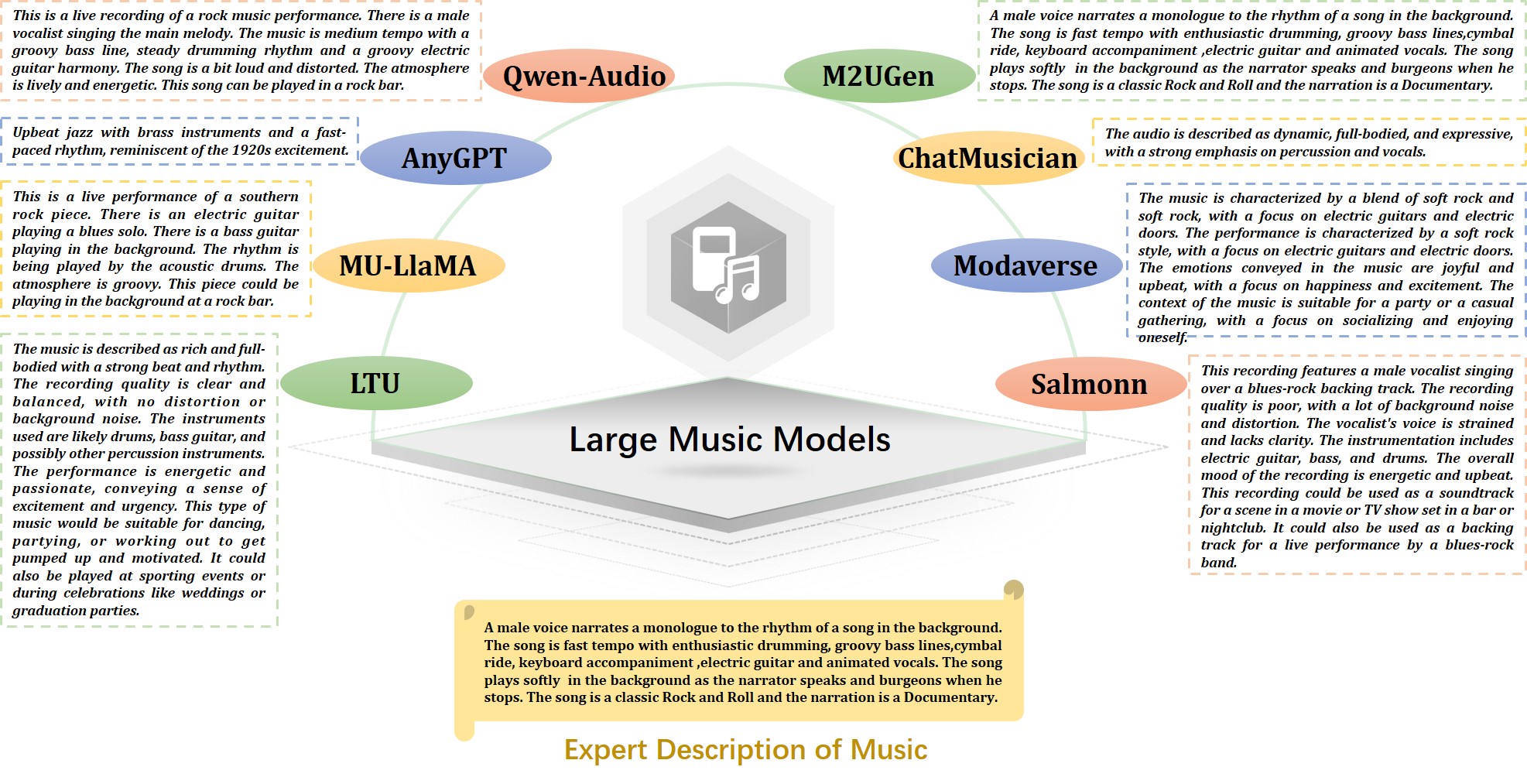}
    \caption{\ We randomly selected a music sample as well as the corresponding caption as the ground truth, and presented the responses provided by each model. A single presentation does not represent the overall performance of the model.}
    \label{fig:response with reference}
\end{figure}

 For music captioning, we used the MusicCaps dataset. In Table \ref{Table: Caption Results}, we have listed the performance of models in understanding complex and advanced tasks in music. In Figure \ref{fig:response with reference}, we show the response of each model for an example music segment. The model with the best overall performance was Qwen-Audio, with the highest Rouge-R1 and RL F1 scores. SALMONN and LTU also demonstrated relatively balanced capabilities in understanding advanced tasks. AnyGPT had the lowest precision, higher recall, but a low F1 score, making it the worst performer overall. The overall performance of other models was not particularly outstanding.

\captionsetup{justification=centering}
\begin{table}[h!]
\centering
\caption{\ Model performance on MusicCaps dataset. The best
performance in each column is highlighted by  \pmb{BOLD} type.  General models (top section) and music specific models (bottom section) are separated.}
\begin{tabular}{lcccccc}
\toprule
Model & R1\_Precision & R1\_Recall & R1\_Fmeasure & RL\_Precision & RL\_Recall & RL\_Fmeasure \\
\midrule
Qwen-Audio & \pmb{0.349} & \pmb{0.356} & \pmb{0.336} & \pmb{0.249} & \pmb{0.255} & \pmb{0.239} \\
LTU & 0.235 & 0.292 & 0.247 & 0.156 & 0.193 & 0.163 \\
SALMONN & 0.325 & 0.235 & 0.254 & 0.216 & 0.256 & 0.167 \\
ModaVerse & 0.176 & 0.288 & 0.198 & 0.134 & 0.226 & 0.152 \\
AnyGPT & 0.092 & 0.307 & 0.138 & 0.073 & 0.241 & 0.108 \\
\cmidrule(lr){1-7}
MU-LLaMA & 0.241 & 0.297 & 0.243 & 0.168 & 0.211 & 0.169 \\
ChatMusician & 0.335 & 0.184 & 0.214 & 0.237 & 0.125 & 0.147 \\
M²UGen & 0.207 & 0.332 & 0.232 & 0.151 & 0.248 & 0.171 \\
\bottomrule
\end{tabular}
\label{Table: Caption Results}
\end{table}

\subsection{Analysis}
Intuitively, it was anticipated that models specifically designed for music would perform better across various tasks. However, the results showed that general multimodal language models outperformed dedicated music models in music understanding tasks. 

\textbf{Music Classification}

 The high classification accuracy of these two models is due to their extensive task training on large datasets, which significantly enhances their generalization ability. These models also provide good answers when asked different questions.
In contrast, M²UGen and ModaVerse have not addressed the basic music understanding task of music classification and lack training on music datasets with classification labels, resulting in poor performance in music classification tasks. Specifically, M²UGen often produces results in a single category or fails to provide any options, while ModaVerse inaccurately generates results across multiple labels and cannot provide accurate answers.

\textbf{Music Caption}

In general, the ROUGE score is relatively low because the dataset references include advanced musical reasoning, not just descriptions of musical phenomena (e.g., a piece of music suitable for opera house performance). The best-performing model is still Qwen-Audio, due to its multi-task pre-training, which includes captioning and question-answering tasks, and allows it to excel in long-form Q\&A. Another key factor is its supervised fine-tuning, which improves the model's alignment with human intentions.

M²UGen, which performed poorly in classification tasks, shows decent performance in music captioning because it was trained on a large number of music-caption data pairs. However, these data pairs were generated based on MU-LLaMA and lacked fine-tuning for advanced reasoning, limiting M²UGen's performance. The most unsatisfactory performance among all models is from AnyGPT. Although AnyGPT can handle multimodal inputs and outputs and perform related tasks, it cannot use prompts in a personalized manner for specific tasks, and it has higher losses compared to single-modal training. These factors contribute to AnyGPT's poor performance in specific modal tasks.

In summary, we find that models with more input and output modalities, such as AnyGPT and M²UGen, do not perform well in music understanding. This may be due to the increase in modalities without a corresponding increase in the number of parameters, or because the weights of different modalities interfere with each other. Models that perform better generally do not include image modalities, meaning they do not have weights for visual information. Additionally, among models with similar input and output modalities, those that perform poorly often lack diverse training data. Some models focus on a single category of music data, such as ChatMusician, which primarily covers Irish music, leading to bias in the model. Another reason for poor performance is the lack of diversity in music instructions. One important reason why Qwen-Audio and the LTU models perform well is that both cover a variety of tasks during training, enabling them to handle different tasks effectively. These factors are significant contributors to the performance differences among models.

\section{Discussion}
After comparing and studying existing general multimodal large foundation models and specialized music understanding models, we suggest that AI holds immense potential in the field of music understanding. Although our comparative results have demonstrated that AI methods can enable machines to understand and appreciate music in some relatively simple tasks, the development of AI in the music understanding domain still faces significant challenges.

\subsection{Limitation of Large Music Model}
Firstly, there is a scarcity of high-quality, manually annotated datasets. Multimodal large foundation models rely heavily on high-quality text-x(X means a certain modality) modality sample pairs for alignment training\cite{yin2023survey,zeng2024matters}. Music, being a complex and artistic modality\cite{tarasti2004music,jensen2022media}, requires understanding not only its components like instruments and harmony but also the human responses such as perception and emotion embedded in it. Creating such annotated datasets necessitates professional music experts to evaluate and label the music, which is a daunting task. However, these datasets are crucial for helping machines\cite{budach2022effects} understand music and the human responses it contains, ultimately enabling more complex and advanced music understanding.

Secondly, existing music understanding and generation models tend to focus more on generative tasks \cite{wang2024review}, lacking sufficient training in music understanding tasks. The interference between music understanding and music generation tasks has not yet been fully addressed. The music generation module of the model may interfere with the music understanding module, especially in models with multi-modal inputs and outputs\cite{wang2020makes,neverova2015moddrop,peng2022balanced}. These models may perform worse than those with fewer modalities, as demonstrated in our experiments.
In addition, the model lacks training for simple music comprehension tasks, which results in poor generalization ability of the model, such as M²UGen.

Additionally, the language models used in existing music understanding are typically either directly adopted or only slightly fine-tuned from existing language models. These models inherently lack the specialized musical knowledge that professional musicians possess, and their music reasoning behavior is based on the corpus designed during the training phase rather than the rules of music itself. Since the language model is a core component in music understanding, its lack of musical expertise makes it difficult to provide expert-level music understanding assessments. Insufficient richness and specialization of the musical knowledge within the language model, causing it to provide unprofessional or unrelated answers and creating significant discrepancies between the model's music understanding and actual human experience. To better understand music, the core language model must undergo more domain-specific fine-tuning\cite{gu2021domain,tinn2023fine,susnjak2024automating}. Or, expert-level language models specifically designed for music must be developed, incorporating more specialized musical knowledge into the language model.

Finally, many models use music data in formats such as .flac, .wav, and .mp3 for music-text modality alignment. These formats capture the physical phenomena of music rather than its ``language" as midi does. When processing music inputs, these models treat music as a generic audio modality, resulting in audio recognition outcomes only. This means that these models miss the essential encoded information of the music itself, making it difficult to grasp the ``essence" of the music. Consequently, these models face challenges in understanding the rules of music comprehensively.

\subsection{ The Future Works of Large Music Model}

To develop specialized models in the music understanding domain, we can use music datasets in various formats for model training. Using .flac and .wav format music datasets allows the model to understand the physical phenomena of music, similar to the development of audio models. Using .midi format music datasets, which store musical performance instructions including notes, dynamics, rhythm, and tempo, enables the model to learn based on the rules of music. Rhythm features can be extracted from the timing of notes, chord features from the combination of notes, and dynamic features from changes in note dynamics. Developing large music models based on the .midi format is a crucial step in the current stage of large foundation model development. By simultaneously using .flac, .wav, and .midi format music datasets, we can incorporate both the physical and semantic information of music, allowing the model to extract musical features from multiple dimensions\cite{cataltepe2007music,loy1985musicians,tzanetakis2002musical,pan1995tutorial}. This approach will aid in developing more intelligent and accurate music understanding models.

As for modeling training, GPT-3.5's significance as a groundbreaking large model offers valuable insights. In its training process, there are two crucial stages: the supervised fine-tuning stage\cite{gunel2020supervised} and the reward modeling stage\cite{ouyang2022training}.
During the supervised fine-tuning stage, although the data volume is low, the quality is high. The training data mainly consists of artificially generated or selected high-quality datasets, with a data volume several orders of magnitude smaller than that of the pre-training stage. Models fine-tuned in this manner can better handle music-related tasks\cite{taori2023stanford,gao2023llama}. This requires collaboration among cross-disciplinary researchers to provide the model with high-quality music datasets that have been carefully selected or manually annotated, enabling the model to extract prominent musical features and learn specialized musical knowledge during the weight update process in the training stage.
The subsequent reward modeling stage involves manually scoring the model's music understanding and ranking each response. These scoring data are then used to further refine the model. Based on these scores, a loss function (LOSS) is designed, and by continuously minimizing the LOSS, the model is adjusted to meet the desired standards, thereby acquiring the rich knowledge base of music experts. This method is called Reinforcement Learning from Human Feedback(RLHF)\cite{bai2022training,sun2023aligning,dai2023safe}. Although these stages are highly dependent on and consume significant human resources\cite{casper2023open}, models trained through these processes achieve excellent results, making their responses highly professional. This is a crucial element in the future development of large music models and will determine the professionalism and quality of the generated outcomes.

Finally, music plays a very important role in emotional regulation\cite{cook2019music}, education\cite{holland2013artificial}, therapy\cite{center2005music}, and emotional expression.For example, music is one of the most important associative forms with human emotional expression\cite{schafer2013psychological}. Conceptually, musical emotion is considered as a challenging-to-quantify expression of human emotions\cite{juslin2001music}, having undergone rich transformations throughout the evolution of music. Currently, the level of intelligence in music understanding is relatively low, primarily based on signal analysis perspectives, lacking the integration of human music emotion recognition systems and anthropomorphic composition thinking\cite{pietikainen2022challenges}. Music emotion recognition is a complex reasoning task in music understanding and a significant challenge for current large foundation models.
To equip models with music emotion recognition capabilities, a large volume of music datasets with emotion labels can be incorporated during the training phase. At present, music datasets with emotion annotations are relatively abundant. In addition to inputting these labeled datasets, supervised fine-tuning\cite{prottasha2022transfer} with a focus on musical emotions can further enhance the model's music understanding performance\cite{feng2021rethinking}.This is a feasible functionality that could be realized in the near term.
In the future, under the umbrella of affective computing\cite{zhao2022review}, achieving intelligent combination and synergy of music robots will be a significant highlight in the development of intelligent music models.

\section{Conclusion}
This article summarizes the current state of development in the field of music understanding using foundation models. We introduce relevant music datasets, subjective evaluation metrics, and objective evaluation metrics. We list several large foundation models related to music understanding and test their performance on basic and advanced music understanding tasks, providing readers with an intuitive understanding of the current performance of large music understanding models. We analyze and summarize the reasons that might cause performance differences in these models. We also discuss the shortcomings and limitations of large foundation models in the current stage of music understanding development from the perspectives of data, language model modules, and music-text modality fusion. We also point out several feasible future research directions, such as starting from the music itself, high-quality supervised fine-tuning, and incorporating human emotional knowledge. We hope this review can provide researchers in the field of music understanding with new insights.


\section*{Acknowledgement}

This work was supported by the National Natural Science Foundation of China under Grant 62131009, 62107029, U23A20335, 62476222, 61936007, 61976131, and the Shaanxi
Province Department of Science and Technology under Grant
2024SF-YBXM-064.

\bibliographystyle{IEEEtranN}  
\bibliography{references.bib} 

\end{document}